%% file: umeda.tex
\documentstyle[epsf]{aipproc}
\input pah0.tex

\newcommand\etal{{ et al. }}
\def\mdot{$\dot M$}
\def\msy{$M_\odot$ yr$^{-1}$}
\def\kms{km s$^{-1}$}
\def\e#1{$\times$ $10^{#1}$ }
\def\ee#1{$10^{#1}$ }
\def\lsim{\mathrel{\rlap{\lower 4pt \hbox{\hskip 1pt $\sim$}}\raise 1pt \hbox
        {$<$}}}
\def\gsim{\mathrel{\rlap{\lower 4pt \hbox{\hskip 1pt $\sim$}}\raise 1pt \hbox
        {$>$}}}
\def\ltsim{\mathrel{\rlap{\lower 4pt \hbox{\hskip 1pt $\sim$}}\raise 1pt \hbox
        {$<$}}}
\def\gtsim{\mathrel{\rlap{\lower 4pt \hbox{\hskip 1pt $\sim$}}\raise 1pt \hbox
        {$>$}}}
\def\apj{ApJ}
\def\aap{A\&A}
\def\aj{AJ}
\def\mnras{MNRAS}
\def\araa{ARAA}
\def\pasp{PASP}
\def\apjs{ApJS}
\def\nat{Nature}

\begin{document}

\title{Type Ia Supernovae: Progenitors and Evolution with Redshift}

\author{Ken'ichi Nomoto$^1$, Hideyuki Umeda$^1$, Chiaki Kobayashi$^1$,\\
Izumi Hachisu$^2$, Mariko Kato$^3$, \& Takuji Tsujimoto$^4$ }

\address{$^1$Department of Astronomy and Research Center for the Early
Universe, University of Tokyo\\
$^2$Department of Earth Science and Astronomy, College of Arts and
Sciences, University of Tokyo\\
$^3$Department of Astronomy, Keio University, Hiyoshi, Yokohama,
Japan\\
$^4$National Astronomical Observatory, Mitaka, Japan\\ 
\bigskip
To be published in the Proceedings of the 10th Annual October
Astrophysics Conference\\ in Maryland (11 - 13 October 1999) 
 "COSMIC EXPLOSIONS!",\\  ed. S.S. Holt and W.W. Zhang 
(American Institute of Physics)}

%\lefthead{LEFT head}
%\righthead{RIGHT head}
\maketitle

\vspace*{-10mm}

\begin{abstract}

Relatively uniform light curves and spectral evolution of Type Ia
supernovae (SNe Ia) have led to the use of SNe Ia as a ``standard
candle'' to determine cosmological parameters.  
Whether a statistically significant value of the cosmological constant
can be obtained depends on whether the peak luminosities of SNe Ia are
sufficiently free from the effects of cosmic and galactic evolutions.

Here we first review the single degenerate scenario for the
Chandrasekhar mass white dwarf (WD) models of SNe Ia.  We identify the
progenitor's evolution and population with two channels: (1) the WD+RG
(red-giant) and (2) the WD+MS (near main-sequence He-rich star)
channels.  In these channels, the strong wind from accreting WDs
plays a key role, which yields important age and metallicity
effects on the evolution.

We then address the questions whether the nature of SNe Ia depends
systematically on environmental properties such as metallicity and age
of the progenitor system and whether significant evolutionary effects
exist.  We suggest that the variation of the carbon mass fraction
$X$(C) in the C+O WD (or the variation of the initial WD mass) causes
the diversity of the brightness of SNe Ia.  This model can explain 
the observed dependences of SNe Ia brightness on the galaxy types 
and the distance from the galactic center.

Finally, applying the metallicity effect on the evolution of SN Ia
progenitors, we make a prediction of the cosmic supernova rate history
as a composite of the supernova rates in different types of galaxies.

\end{abstract}

\vspace*{-3mm}
\section{Introduction}
\vspace*{-1mm}

Type Ia supernovae (SNe Ia) are good distance indicators, and provide
a promising tool for determining cosmological parameters (e.g.,
\cite{bra98}).  SNe Ia have been discovered up to $z \sim 1.32$
\cite{gil99}.  Both the Supernova Cosmology Project \cite{per97,per99}
and the High-z Supernova Search Team \cite{gar98,rie98} have suggested
a statistically significant value for the cosmological constant.

However, SNe Ia are not perfect standard candles, but show some
intrinsic variations in brightness.  When determining the absolute
peak luminosity of high-redshift SNe Ia, therefore, these analyses
have taken advantage of the empirical relation existing between the
peak brightness and the light curve shape (LCS).  Since this relation
has been obtained from nearby SNe Ia only \cite{phi93,ham95,rie95}, 
it is important to examine whether it depends
systematically on environmental properties such as metallicity and age
of the progenitor system.

High-redshift supernovae present us very useful information, not only
to determine cosmological parameters but also to put constraints on
the star formation history in the universe.  They have given the SN Ia
rate at $z \sim 0.5$ \cite{pai99} but will provide the SN Ia rate
history over $0<z<1$.  With the Next Generation Space Telescope, both
SNe Ia and SNe II will be observed through $z \sim 4$.  It is useful
to provide a prediction of cosmic supernova rates to constrain the age
and metallicity effects of the SN Ia progenitors.

SNe Ia have been widely believed to be a thermonuclear explosion of a
mass-accreting white dwarf (WD) (e.g., \cite{nom97a} for a
review).  However, the immediate progenitor binary systems have not
been clearly identified yet \cite{bra95}.  In order to address the
above questions regarding the nature of high-redshift SNe Ia, we need
to identify the progenitors systems and examine the ``evolutionary''
effects (or environmental effects) on those systems.

In \S2, we summarize the progenitors' evolution where the strong wind
from accreting WDs plays a key role
\cite{hac96,hac99a,hac99b}.
In \S3, we address the issue of whether a difference in
the environmental properties is at the basis of the observed range of
peak brightness \cite{ume99b}.  In \S4, we make a prediction of
the cosmic supernova rate history as a composite of the different
types of galaxies \cite{kob00}.

\vspace*{-1mm}
\section{Evolution of progenitor systems}
\vspace*{-1mm}

There exist two models proposed as progenitors of SNe Ia: 1) the
Chandrasekhar mass model, in which a mass-accreting carbon-oxygen
(C+O) WD grows in mass up to the critical mass $M_{\rm Ia} \simeq
1.37-1.38 M_\odot$ near the Chandrasekhar mass and explodes as an SN
Ia (e.g., \cite{nom84,nom94}), and 2) the sub-Chandrasekhar mass
model, in which an accreted layer of helium atop a C+O WD ignites
off-center for a WD mass well below the Chandrasekhar mass
(e.g., \cite{arn96}). 
The early time spectra of the majority of SNe Ia are in
excellent agreement with the synthetic spectra of the Chandrasekhar
mass models, while the spectra of the sub-Chandrasekhar mass models
are too blue to be comparable with the observations \cite{hof96,nug97}.

     For the evolution of accreting WDs toward the Chandrasekhar mass,
two scenarios have been proposed: 1) a double degenerate (DD)
scenario, i.e., merging of double C+O WDs with a combined mass
surpassing the Chandrasekhar mass limit \cite{ibe84,web84},
and 2) a single degenerate (SD) scenario, i.e., accretion of
hydrogen-rich matter via mass transfer from a binary companion
(e.g., \cite{nom82,nom94}).  The issue of DD vs. SD is still 
debated (e.g., \cite{bra95}), although theoretical modeling has
indicated that the merging of WDs leads to the accretion-induced
collapse rather than SN Ia explosion 
\cite{sai85,sai98,seg97}.

     In the SD Chandrasekhar mass model for SNe Ia, a WD explodes as a
SN Ia only when its rate of the mass accretion ($\dot M$) is in a
certain narrow range (e.g., \cite{nom82,nom91}).  In
particular, if $\dot M$ exceeds the critical rate $\dot M_{\rm b}$, 
the accreted matter extends to form a common envelope
\cite{nom79}.  This difficulty has been overcome by the WD
wind model (see below).  For the actual binary systems which grow the
WD mass ($M_{\rm WD}$) to $M_{\rm Ia}$, the following two systems are
appropriate.  One is a system consisting of a mass-accreting WD and a
lobe-filling, more massive, slightly evolved main-sequence or
sub-giant star (hereafter ``WD+MS system'').  The other system
consists of a WD and a lobe-filling, less massive, red-giant
(hereafter ``WD+RG system'').

\vspace*{-5mm}
\subsection {White dwarf winds}
\vspace*{-5mm}

Optically thick WD winds are driven when the accretion rate $\dot M$
exceeds the critical rate $\dot M_{\rm b}$.  Here $\dot M_{\rm b}$ is
the rate at which steady burning can process the accreted hydrogen
into helium as 
$\dot M_{\rm b} \approx 0.75 \times10^{-6} \left({M_{\rm WD} \over {M_\odot}}
 - 0.40\right) M_\odot {\rm ~yr}^{-1}$.

With such a rapid accretion, the WD envelope expands to $R_{\rm ph}
\sim 0.1 R_\odot$ and the photospheric temperature decreases below
$\log T_{\rm ph} \sim 5.5$.  Around this temperature, the shoulder of
the strong peak of OPAL Fe opacity \cite{igl93} drives the
radiation-driven wind \cite{hac96,hac99b}.  
The ratio of $v_{\rm ph}/v_{\rm esc}$ 
between the photospheric velocity and the escape velocity 
at the photosphere depends on the mass transfer rate and $M_{\rm WD}$.
(see Fig.6 in \cite{hac99b}).  
We call the wind {\sl strong} when $v_{\rm ph}>v_{\rm esc}$. 
When the wind is strong, $v_{\rm ph} \sim 1000$ km
s$^{-1}$ being much faster than the orbital velocity.

If the wind is sufficiently strong, the WD can avoid the formation of
a common envelope and steady hydrogen burning increases its mass
continuously at a rate $\dot M_{\rm b}$ by blowing the extra mass away
in a wind.  When the mass transfer rate decreases below this critical
value, optically thick winds stop.  If the mass transfer rate further
decreases below $\sim$ 0.5 $\dot M_{\rm b}$, hydrogen shell burning
becomes unstable to trigger very weak shell flashes but still burns
a large fraction of accreted hydrogen.

     The steady hydrogen shell burning converts hydrogen into helium
atop the C+O core and increases the mass of the helium layer
gradually.  When its mass reaches a certain value, weak helium shell
flashes occur.  Then a part of the envelope mass is blown off but a
large fraction of He can be burned to C+O \cite{kat99h} to increase
the WD mass.  In this way, strong winds from the accreting WD play a
key role to increase the WD mass to $M_{\rm Ia}$.

\vspace*{-3mm}
\subsection{WD+RG system} 
\vspace*{-3mm}

This is a symbiotic binary system consisting of a WD and
a low mass red-giant (RG).  A full evolutionary path of the WD+RG
system from the zero age main-sequence stage to the SN
Ia explosion is described in \cite{hac99b,tut77}.
The occurrence frequency of SNe Ia through this channel is much
larger than the earlier scenario, because of the following two
evolutionary processes, which have not considered before.

(1) Because of the AGB wind, the WD + RG close binary can
form from a wide binary even with such a large initial separation as
$a_i \lsim 40,000 R_\odot$.  Our earlier estimate \cite{hac96} is
constrained by $a_i \lsim 1,500 R_\odot$.

(2) When the RG fills its inner critical Roche lobe, the WD undergoes
rapid mass accretion and blows a strong optically thick wind.  Our
earlier analysis has shown that the mass transfer is stabilized by
this wind only when the mass ratio of RG/WD is smaller than 1.15.  Our
new finding is that the WD wind can strip mass from the RG envelope,
which could be efficient enough to stabilize the mass transfer even if
the RG/WD mass ratio exceeds 1.15.  If this mass-stripping effect is
strong enough, though its efficiency $\eta_{\rm eff}$ is subject to
uncertainties, the symbiotic channel can produce SNe Ia for a much
(ten times or more) wider range of the binary parameters than our
earlier estimation.

With the above two new effects (1) and (2), the WD+RG (symbiotic)
channel can account for the inferred rate of SNe Ia in our Galaxy.
The immediate progenitor binaries in this symbiotic channel to SNe Ia
may be observed as symbiotic stars, luminous supersoft X-ray sources,
or recurrent novae like T CrB or RS Oph, depending on the wind status.

\vspace*{-3mm}
\subsection{WD+MS system} 
\vspace*{-3mm}

In this scenario, a C+O WD is originated, not from an AGB star with a
C+O core, but from a red-giant star with a helium core of $\sim
0.8-2.0 M_\odot$.  The helium star, which is formed after the first
common envelope evolution, evolves to form a C+O WD of $\sim 0.8-1.1
M_\odot$ with transferring a part of the helium envelope onto the
secondary main-sequence star.  A full evolutionary path of the WD+MS
system from the zero age main-sequence stage to the SN Ia explosion is
described in \cite{hac99a}.

This evolutionary path provides a much wider channel to SNe Ia than
previous scenarios.  A part of the progenitor systems are identified
as the luminous supersoft X-ray sources \cite{heu92}
during steady H-burning (but without wind to avoid extinction), or the
recurrent novae like U Sco if H-burning is weakly unstable.  Actually
these objects are characterized by the accretion of helium-rich
matter.

\vspace*{-3mm}
\subsection{Realization frequency}
\vspace*{-3mm}

%%% fig 1
\begin{figure}
\centerline{\psfig{figure=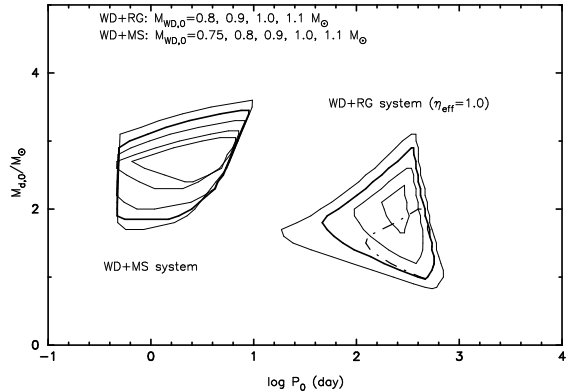,width=8cm}}
\caption[h]{\label{ztotreg100}
The region to produce SNe Ia in the $\log P_0 - M_{\rm d,0}$ plane for
five initial WD masses of $0.75 M_\odot$, $0.8 M_\odot$, $0.9
M_\odot$, $1.0 M_\odot$ (heavy solid line), and $1.1 M_\odot$.  The
region of $M_{\rm WD,0}= 0.7 M_\odot$ almost vanishes for both the
WD+MS and WD+RG systems, and the region of $M_{\rm WD,0}= 0.75
M_\odot$ vanishes for the WD+RG system.  Here, we assume the stripping
efficiency of $\eta_{\rm eff}=1$.  For comparison, we show only the
region of $M_{\rm WD,0}= 1.0 M_\odot$ for a much lower efficiency of
$\eta_{\rm eff}=0.3$ by a dash-dotted line.}
\end{figure}

     For an immediate progenitor system WD+RG of SNe Ia, we consider a
close binary initially consisting of a C+O WD with $M_{\rm WD,0}=
0.6-1.2 M_{\odot}$ and a low-mass red-giant star with $M_{\rm RG,0}=
0.7-3.0 M_{\odot}$ having a helium core of $M_{\rm He,0}= 0.2-0.46
M_{\odot}$. The initial state of these immediate progenitors
is specified by three parameters, i.e., $M_{\rm WD,0}$, $M_{\rm RG,0}
= M_{\rm d,0}$, and the initial orbital period $P_0$ ($M_{\rm He,0}$
is determined if $P_0$ is given).

     We follow binary evolutions of these systems and obtain the
parameter range(s) which can produce an SN Ia.  In Figure
\ref{ztotreg100}, the region enclosed by the thin solid line produces
SNe Ia for several cases of the initial WD mass, $M_{\rm WD,0} =$ 0.75
- 1.1 $M_\odot$.  For smaller $M_{\rm WD,0}$, the wind is weaker, so
that the SN Ia region is smaller. The regions of $M_{\rm WD,0} = 0.6
M_\odot$ and $0.7 M_\odot$ vanish for both the WD+MS and WD+RG
systems.  

     In the outside of this region, the outcome of the evolution at
the end of the calculations is not an SN Ia but one of the followings:
(i)
  Formation of a common envelope for too large $M_{\rm d}$ or $P_0
\sim$ day, where the mass transfer is unstable at the beginning of
mass transfer.
(ii)
  Novae or strong hydrogen shell flash for too small $M_{\rm d,0}$,
where the mass transfer rate becomes below \ee{-7} \msy.
(iii)
  Helium core flash of the red giant component for too long $P_0$,
where a central helium core flash ignites, i.e., the helium core mass
of the red-giant reaches $0.46 M_\odot$.
(iv)
  Accretion-induced collapse for $M_{\rm WD,0} > 1.2 M_\odot$, where
the central density of the WD reaches $\sim 10^{10}$ g cm$^{-3}$
before heating wave from the hydrogen burning layer reaches the
center.  As a result, the WD undergoes collapse due to electron
capture without exploding as an SN Ia \cite{nom91}.

     It is clear that the new region of the WD+RG system is not
limited by the condition of $q < 1.15$, thus being ten times or more
wider than the region of \cite{hac96}'s model (depending on the the stripping
efficiency of $\eta_{\rm eff}$).  

     The WD+MS progenitor system can also be specified by three
initial parameters: the initial C+O WD mass $M_{\rm WD,0}$, the mass
donor's initial mass $M_{\rm d,0}$, and the orbital period $P_0$.  For
$M_{\rm WD,0} = 1.0 M_\odot$, the region producing an SN Ia is bounded
by $M_{\rm d,0}= 1.8-3.2 M_\odot$ and $P_0= 0.5-5$ d as shown by the
solid line in Figure \ref{ztotreg100}.  The upper and lower bounds are
respectively determined by the common envelope formation (i) and
nova-like explosions (ii) as above.  The left and right bounds are
determined by the minimum and maximum radii during the main sequence
of the donor star \cite{hac99a}.

     We estimate the rate of SNe Ia originating from these channels in
our Galaxy by using equation (1) of \cite{ibe84}.  The
realization frequencies of SNe Ia through the WD+RG and WD+MS channels
are estimated as $\sim$ 0.0017 yr$^{-1}$ (WD+RG) and $\sim$ 0.001
yr$^{-1}$ (WD+MS), respectively.  The total SN Ia rate of the
WD+MS/WD+RG systems becomes $\sim$ 0.003 yr$^{-1}$, which is close
enough to the inferred rate of our Galaxy.

\vspace*{-3mm}
\subsection{Low metallicity inhibition of type Ia supernovae}
\vspace*{-3mm}

%%% fig 2, 3
\begin{figure}
\begin{minipage}[t]{0.47\textwidth}
\centerline{\psfig{figure=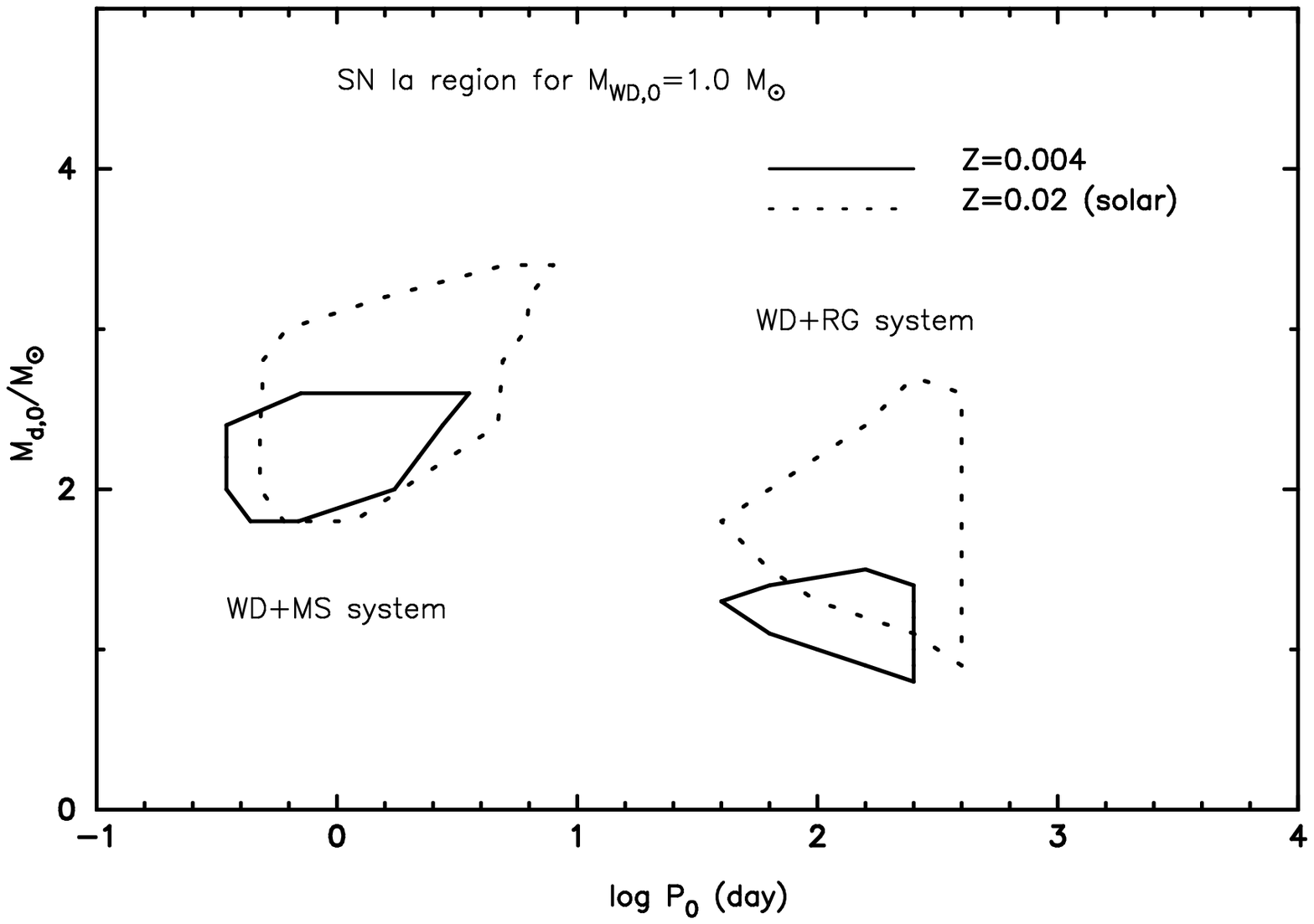,width=7.0cm}}
\caption{\label{koba_fig2}
The regions of SNe Ia is plotted in the initial orbital period vs. the
initial companion mass diagram for the initial WD mass of $M_{\rm
WD,0}=1.0 M_\odot$.  The dashed and solid lines represent the cases of
solar abundance ($Z=0.02$) and much lower metallicity of $Z=0.004$,
respectively. The left and the right regions correspond to the WD+MS
and the WD+RG systems, respectively.}
\end{minipage}
\hspace*{\fill}
\begin{minipage}[t]{0.47\textwidth}
\centerline{\psfig{figure=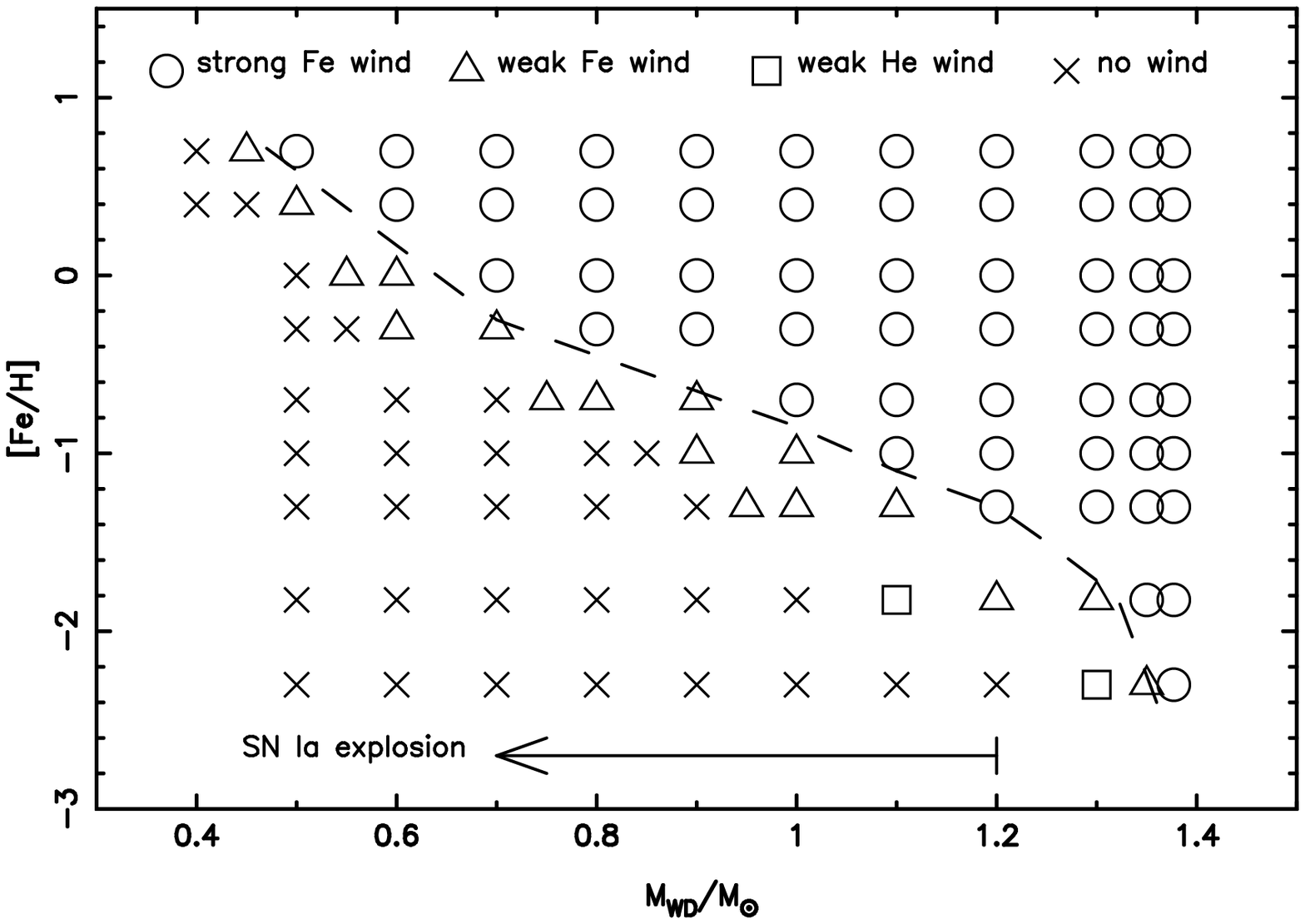,width=7.0cm}}
\caption{\label{koba_fig1}
WD mass vs. metallicity diagram showing the metallicity dependence of
optically thick winds. We regard the wind as ``strong'' 
if $v_{\rm w}>v_{\rm esc}$ but ``weak'' if $v_{\rm w}<v_{\rm esc}$. 
The term of
``He'' or ``Fe'' wind denotes that the wind is accelerated by the peak
of iron lines near $\log T ({\rm K}) \sim 5.2$ or of helium lines near
$\log T ({\rm K}) \sim 4.6$.  The dashed line indicates the
demarcation between the ``strong'' wind and the ``weak'' wind.}
\end{minipage}
\end{figure}

The optically thick winds are driven by a strong peak of OPAL opacity
at $\log T ({\rm K}) \sim 5.2$ (e.g., \cite{igl93}). Since the opacity
peak is due to iron lines, the wind velocity $v_{\rm w}$ depends on
the iron abundance [Fe/H] (\cite{kob98,hac00}), i.e., $v_{\rm w}$ is
higher for larger [Fe/H].  The metallicity effect on SNe Ia is clearly
demonstrated by the size of the regions to produce SNe Ia in the
diagram of the initial orbital period versus initial mass of the
companion star (see Fig. \ref{koba_fig2}).  The SN Ia regions are much
smaller for lower metallicity because the wind becomes weaker.  

\medskip

The wind velocity depends also on the luminosity $L$ of the WD.  The
more massive WD has a higher $L$, thus blowing higher velocity winds
(\cite{hac99b}).  In order for the wind velocity to exceed the escape
velocity of the WD near the photosphere, the WD mass should be larger
than a certain critical mass for a given [Fe/H].  This implies that
the initial mass of the WD $M_{\rm WD, 0}$ should already exceed that
critical mass in order for the WD mass to grow to the Ch mass.  This
critical mass is larger for smaller [Fe/H], reaching $1.1 M_\odot$ for
[Fe/H] $=-1.1$ (Fig. \ref{koba_fig1}).  Here we should note that the
relative number of WDs with $M_{\rm WD,0} \gtsim 1.1 M_\odot$ is quite
small in close binary systems (\cite{ume99a}).  And for $M_{\rm WD,0}
\gtsim 1.2 M_\odot$, the accretion leads to collapse rather than SNe
Ia (\cite{nom91}).  Therefore, no SN Ia occurs at [Fe/H] $\le -1.1$ in
our model.

It is possible to test the metallicity effects on SNe Ia with the
chemical evolution of galaxies.

In the one-zone uniform model for the chemical evolution of the solar
neighborhood, the heavy elements in the metal-poor stars originate
from the mixture of the SN II ejecta of various progenitor masses.
The abundances averaged over the progenitor masses of SNe II predicts
[O/Fe] $\sim 0.45$ (e.g., \cite{tsu95,nom97c}).  Later SNe Ia start
ejecting mostly Fe, so that [O/Fe] decreases to $\sim 0$ around [Fe/H]
$\sim 0$.  The low-metallicity inhibition of SNe Ia predicts that the
decrease in [O/Fe] starts at [Fe/H] $\sim -1$.  Such an evolution of
[O/Fe] well explains the observations (\cite{kob98}).

However, we should note that some anomalous stars have [O/Fe] $\sim$ 0
at [Fe/H] $\ltsim -1$.  The presence of such stars, however, is not in
conflict with our SNe Ia models, but can be understood as follows: The
formation of such anomalous stars (and the diversity of [O/Fe] in
general) indicates that the interstellar materials were not uniformly
mixed but contaminated by only a few SNe II (or even single SN II)
ejecta.  This is because the timescale of mixing was longer than the
time difference between the supernova event and the next generation
star formation.  The iron and oxygen abundances produced by a single
SN II vary depending on the mass, energy, mass cut, and metallicity of
the progenitor.  Relatively smaller mass SNe II ($13-15 M_\odot$) and
higher explosion energies tend to produce [O/Fe] $\sim 0$
(\cite{nom97c,ume00}).  Those metal poor stars with [O/Fe] $\sim 0$
may be born from the interstellar medium polluted by such SNe II.

The metallicity effect on SNe Ia can also be checked with the
metallicity of the host galaxies of nearby SNe Ia.  There has been no
evidence that SNe Ia have occurred in galaxies with a metallicity of
[Fe/H] $\ltsim -1$, although host galaxies are detected only for one
third of SNe Ia and the estimated metallicities of host galaxies are
uncertain.  Three SNe Ia are observed in low-metallicity dwarf
galaxies; SN1895B and SN1972E in NGC 5253, and SN1937C in IC 4182.
Metallicities of these galaxies are estimated to be [O/H] $=-0.25$ and
$-0.35$, respectively \cite{koc97}.  If [O/Fe] $\sim 0$ as in the
Magellanic Clouds, [Fe/H]$\sim -0.25$ and $-0.35$ which are not so
small.  Even if these galaxies have extremely SN II like abundance as
[O/Fe] $\sim 0.45$, [Fe/H] $\sim -0.7$ and $-0.8$ (being higher than
$-1$), respectively.  Since these host galaxies are blue ($B-V=0.44$
for NGC 5253 and $B-V=0.37$ for IC 4182 according to RC3 catalog), the
MS+WD systems are dominant progenitors for the present SNe Ia.  The
rate of SNe Ia originated from the MS+WD systems is not so sensitive
to the metallicity as far as [Fe/H] $> -1$ (\cite{hac00}).  Even if
[Fe/H] $\sim -0.7$ in such blue galaxies, therefore, the SN Ia rate is
predicted to be similar to those in more metal-rich galaxies.

% ------------------------------------------------------

\vspace*{-1mm}
\section{The origin of diversity of SNe Ia and environmental effects}
\vspace*{-1mm}

 There are some observational indications that SNe Ia are affected by
their environment. The most luminous SNe Ia seem to occur only in
spiral galaxies, while both spiral and elliptical galaxies are hosts
for dimmer SNe Ia. Thus the mean peak brightness is dimmer in
ellipticals than in spiral galaxies \cite{ham96}. The SNe Ia
rate per unit luminosity at the present epoch is almost twice as high
in spirals as in ellipticals \cite{cap97}.  Moreover, 
\cite{wan97,rie99} found that the variation of the
peak brightness for SNe located in the outer regions in galaxies is
smaller.

\cite{hof98,hof99} examined how the initial composition of
the WD (metallicity and the C/O ratio) affects the observed properties
of SNe Ia.  \cite{ume99a} obtained the C/O ratio as a function
of the main-sequence mass and metallicity of the WD progenitors.
\cite{ume99b} suggested that the variation of the C/O ratio is
the main cause of the variation of SNe Ia brightness, with larger C/O
ratio yielding brighter SNe Ia.  We will show that the C/O ratio
depends indeed on environmental properties, such as the metallicity
and age of the companion of the WD, and that our model can explain
most of the observational trends discussed above. We then make some
predictions about the brightness of SN Ia at higher redshift.

\vspace*{-3mm}
\subsection{C/O ratio in WD progenitors}
\vspace*{-3mm}

%%% fig 4
\begin{figure} 
\centerline{\psfig{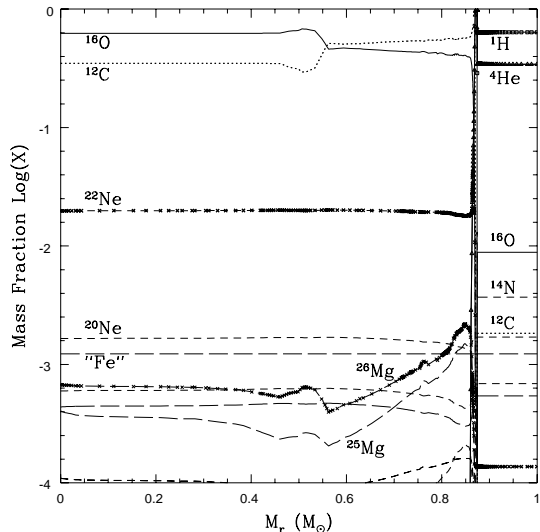}}
\caption[h]{
Abundances in mass fraction in the inner core of the 6 $M_\odot$ star
for $Y$ = 0.2775 and $Z$ = 0.02 at the end of the second dredge-up.  
\label{umefig1}}
\end{figure}

 In this section we discuss how the C/O ratio in the WD depends on the
metallicity and age of the binary system.  The C/O ratio in C+O WDs
depends primarily on the main-sequence mass of the WD progenitor and
on metallicity. 

 We calculated the evolution of intermediate-mass ($3-9 M_\odot$)
stars for metallicity $Z$=0.001 -- 0.03.  In the ranges of stellar
masses and $Z$ considered in this paper, the most important
metallicity effect is that the radiative opacity is smaller for lower
$Z$. Therefore, a star with lower $Z$ is brighter, thus having a
shorter lifetime than a star with the same mass but higher $Z$.  In
this sense, the effect of reducing metallicity for these stars is
almost equivalent to increasing a stellar mass.

 For stars with larger masses and/or smaller $Z$, the luminosity is
higher at the same evolutionary phase.  With a higher nuclear energy
generation rate, these stars have larger convective cores during H and
He burning, thus forming larger He and C-O cores.

 As seen in Figure \ref{umefig1}, the central part of these stars
is oxygen-rich. The C/O ratio is nearly constant in the innermost
region, which was a convective core during He burning.  Outside this
homogeneous region, where the C-O layer grows due to He shell burning,
the C/O ratio increases up to C/O $\gsim 1$; thus the oxygen-rich core
is surrounded by a shell with C/O $\gsim$ 1. In fact this is a generic
feature in all models we calculated. The C/O ratio in the shell is C/O
$\simeq$ 1 for the star as massive as $\sim 7 M_\odot$, and C/O $>1$
for less massive stars.

\begin{figure} 
\centerline{\psfig{figure=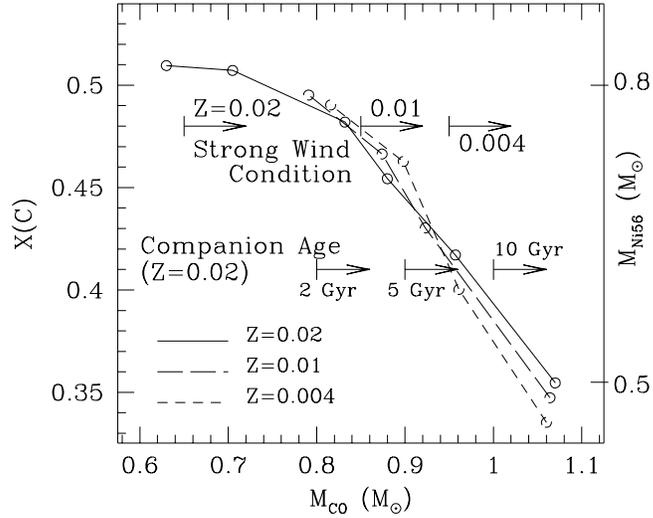,width=9cm}}
\caption[h]{
The total $^{12}$C mass fraction included in the convective core of
mass, $M=1.14M_\odot$, just before the SN Ia explosion as a function
of the C+O core mass before the onset of mass accretion, 
$M_{\rm CO}$. The lower bounds of $M_{\rm CO}$ obtained from the age
effects and the conditions for strong wind to blow are also shown by
arrows.
\label{umefig2}}
\end{figure}

When a progenitor reaches the critical mass for the SNe Ia explosion,
the central core is convective up to around 1.1 $M_\odot$. Hence the
relevant C/O ratio is between the central value before convective
mixing and the total C/O of the whole WD. Using the results from the
C6 model \cite{nom84}, we assume that the convective region is
1.14 $M_\odot$ and for simplicity, C/O = 1 outside the C-O core at the
end of second dredge-up. Then we obtain the C/O ratio of the inner
part of the SNe Ia progenitors (Fig. \ref{umefig2}).

%%% fig 5

 From this figure we find three interesting trends.  First, while the
central C/O is a complicated function of stellar mass \cite{ume99a},
as shown here the C/O ratio in the core before SNe Ia explosion
is a decreasing monotonic function of mass.  The central C/O ratio at
the end of second dredge-up decreases with mass for $M_{\rm ms} \gsim
5M_\odot$, while the ratio increases with mass for $M_{\rm ms} \gsim
4M_\odot$; however, the convective core mass during He burning is
smaller for a less massive star, and the C/O ratio during shell He
burning is larger for smaller C+O core. Hence, when the C/O ratio is
averaged over 1.1 $M_\odot$ the C/O ratio decreases with mass. Second,
as shown in \cite{ume99a}, although the C/O ratio is a complicated
function of metallicity and mass, the metallicity dependence is
remarkably converged when the ratio is seen as a function of the C+O
core mass ($M_{\rm CO}$) instead of the initial main sequence mass.

According to the evolutionary calculations for 3$-$9 $M_\odot$ stars
by \cite{ume99a}, the C/O ratio and its distribution are
determined in the following evolutionary stages of the close binary.

(1) At the end of central He burning in the 3$-$9 $M_\odot$ primary
star, C/O$<1$ in the convective core. The mass of the core is larger
for more massive stars. 

(2) After central He exhaustion, the outer C+O layer grows via He
shell burning, where C/O$\gsim 1$ \cite{ume99a}.

(3a) If the primary star becomes a red giant (case C evolution;
e.g., \cite{van94}), it then undergoes the second dredge-up,
forming a thin He layer, and enters the AGB phase. The C+O core mass,
$M_{\rm CO}$, at this phase is larger for more massive stars. For a
larger $M_{\rm CO}$ the total carbon mass fraction is smaller. 

(3b) When it enters the AGB phase, the star greatly expands and is
assumed here to undergo Roche lobe overflow (or a super-wind phase)
and to form a C+O WD. Thus the initial mass of the WD, $M_{\rm
WD,0}$, in the close binary at the beginning of mass accretion is
approximately equal to $M_{\rm CO}$.

(4a) If the primary star becomes a He star (case BB evolution), the
second dredge-up in (3a) corresponds to the expansion of the He
envelope.

(4b) The ensuing Roche lobe overflow again leads to a WD of
mass $M_{\rm WD,0}$ = $M_{\rm CO}$.

(5) After the onset of mass accretion, the WD mass grows through
steady H burning and weak He shell flashes, as described in the WD
wind model.  The composition of the growing C+O layer is assumed to be
C/O=1.

(6) The WD grows in mass and ignites carbon when its mass reaches
$M_{\rm Ia} =1.367 M_\odot$, as in the model C6 of \cite{nom84}.
Because of strong electron-degeneracy, carbon burning is
unstable and grows into a deflagration for a central temperature of
$8\times 10^8$ K and a central density of $1.47\times 10^9$ g
cm$^{-3}$.  At this stage, the convective core extends to $M_r =
1.14M_\odot$ and the material is mixed almost uniformly, as in the C6
model.

In Figure \ref{umefig2}, we show the carbon mass fraction $X$(C) in
the convective core of this pre-explosive WD, as a function of
metallicity ($Z$) and initial mass of the WD before the onset of mass
accretion, $M_{\rm CO}$.  Figure \ref{umefig2} reveals that: 1) $X$(C)
is smaller for larger $M_{\rm CO} \simeq M_{\rm WD,0}$.  2) The
dependence of $X$(C) on metallicity is small when plotted against
$M_{\rm CO}$, even though the relation between $M_{\rm CO}$ and the
initial stellar mass depends sensitively on $Z$ \cite{ume99a}.

\vspace*{-3mm}
\subsection{Brightness of SNe Ia and the C/O ratio}
\vspace*{-3mm}

In the Chandrasekhar mass models for SNe Ia, the brightness of SNe Ia
is determined mainly by the mass of $^{56}$Ni synthesized ($M_{\rm
Ni56}$). Observational data suggest that $M_{\rm Ni56}$ for most SNe
Ia lies in the range $M_{\rm Ni56} \sim 0.4 - 0.8 M_\odot$
(e.g., \cite{maz98}). This range of $M_{\rm Ni56}$ can result
from differences in the C/O ratio in the progenitor WD as follows.

 In the deflagration model, a larger C/O ratio leads to the production
of more nuclear energy and buoyancy force, thus leading to a faster
propagation.  The faster propagation of the convective deflagration
wave results in a larger $M_{\rm Ni56}$. For example, a variation of
the propagation speed by 15\% in the W6 -- W8 models results in
$M_{\rm Ni56}$ values ranging between 0.5 and $0.7 M_\odot$ \cite{nom84},
which could explain the observations.

In the delayed detonation model, $M_{\rm Ni56}$ is predominantly
determined by the deflagration-to-detonation-transition (DDT) density
$\rho_{\rm DDT}$, at which the initially subsonic deflagration turns
into a supersonic detonation \cite{kho91}.  As discussed in 
\cite{ume99b}, $\rho_{\rm DDT}$ could be very sensitive to $X$(C),
and a larger $X$(C) is likely to result in a larger $\rho_{\rm DDT}$
and $M_{\rm Ni56}$.

Here we postulate that $M_{\rm Ni56}$ and consequently brightness of a
SN Ia increase as the progenitors' C/O ratio increases (and thus
$M_{\rm WD,0}$ decreases).  As illustrated in Figure \ref{umefig2},
the range of $M_{\rm Ni56} \sim 0.5-0.8 M_\odot$ is the result of an
$X$(C) range $0.35-0.5$, which is the range of $X$(C) values of our
progenitor models.  The $X$(C) -- $M_{\rm Ni56}$ -- $M_{\rm WD,0}$
relation we adopt is still only a working hypothesis, which needs to
be proved from studies of the turbulent flame during explosion
(e.g., \cite{nie95}).

\vspace*{-3mm}
\subsection{Metallicity and age effects}
\vspace*{-3mm}

%%% fig 6
\begin{figure}
\centerline{\psfig{figure=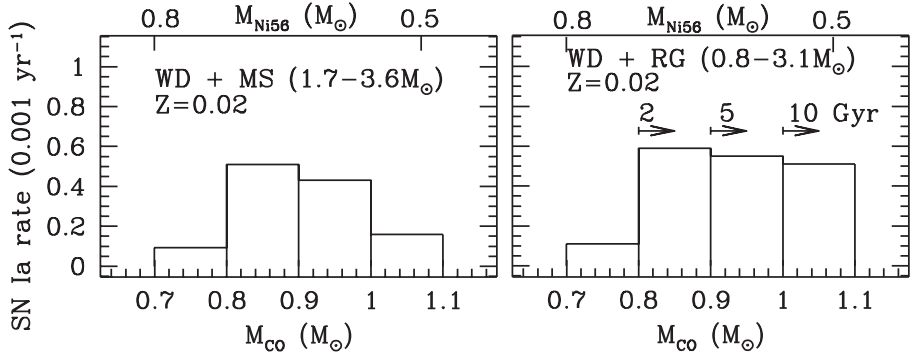,width=12cm}}
\caption[h]{SN Ia frequency for a galaxy of mass $2 \times 10^{11} M_\odot$
as a function of $M_{\rm CO}$ for $Z$=0.02.
For the WD+RG system, constraints from the companion's age are
shown by the arrows. SNe Ia from the WD+MS system occur in spirals
but not in ellipticals because of the age effect. $M_{\rm CO}$ and
$M_{\rm Ni56}$ is assumed to be related as shown here.
\label{umefig3}}
\end{figure}

\vspace*{-1mm}
\subsubsection{Metallicity effects on the minimum $M_{\rm WD,0}$}
\vspace*{-1mm}

As mentioned in \S2.5, $M_{\rm w}$ is the metallicity-dependent
minimum $M_{\rm WD,0}$ for a WD to become an SN Ia ({\sl strong wind
condition} in Fig. \ref{umefig2}).  The upper bound $M_{\rm WD,0}
\simeq 1.07M_\odot$ is imposed by the condition that carbon should not
ignite and is almost independent of metallicity.  As shown in Figure
\ref{umefig2}, the range of $M_{\rm CO} \simeq M_{\rm WD,0}$ can be
converted into a range of $X$(C). From this we find the following
metallicity dependence for $X$(C):

(1) The upper bound of $X$(C), which is determined by the lower limit
on $M_{\rm CO}$ imposed by the metallicity-dependent conditions for a
strong wind, e.g., $X$(C) $\lsim 0.51$, 0.46 and 0.41, for $Z$=0.02,
0.01, and 0.004, respectively.

(2) On the other hand, the lower bound, $X$(C) $\simeq 0.35-0.33$,
does not depend much on $Z$, since it is imposed by the maximum
$M_{\rm CO}$.

(3) Assuming the relation between $M_{\rm Ni56}$ and $ X$(C) given in
Figure \ref{umefig2}, our model predicts the absence of brighter SNe
Ia in lower metallicity environment.

\vspace*{-1mm}
\subsubsection{Age effects on the minimum $M_{\rm WD,0}$} 
\vspace*{-1mm}

In our model, the age of the progenitor system also constrains the
range of $X$(C) in SNe Ia. In the SD scenario, the lifetime of the
binary system is essentially the main-sequence lifetime of the
companion star, which depends on its initial mass $M_2$. \cite{hac99a,hac99b} 
have obtained a constraint on $M_2$ by calculating the evolution
of accreting WDs for a set of initial masses of the WD ($M_{\rm WD,0}
\simeq M_{\rm CO}$) and of the companion ($M_2$), and the initial
binary period ($P_0$). In order for the WD mass to reach $M_{\rm Ia}$,
the donor star should transfer enough material at the appropriate
accretion rates.  The donors of successful cases are divided into two
categories: one is composed of slightly evolved main-sequence stars
with $M_2 \sim 1.7 - 3.6M_\odot$ (for $Z$=0.02), and the other of
red-giant stars with $M_2 \sim 0.8 - 3.1M_\odot$ (for $Z$=0.02)
(Fig. \ref{ztotreg100}).

If the progenitor system is older than 2 Gyr, it should be a system
with a donor star of $M_2 < 1.7 M_\odot$ in the red-giant branch.
Systems with $M_2 > 1.7 M_\odot$ become SNe Ia in a time shorter than
2 Gyr.  Likewise, for a given age of the progenitor system, $M_2$ must
be smaller than a limiting mass. This constraint on $M_2$ can be
translated into the presence of a minimum $M_{\rm CO}$ for a given
age, as follows: For a smaller $M_2$, i.e. for the older system, the
total mass which can be transferred from the donor to the WD is
smaller. In order for $M_{\rm WD}$ to reach $M_{\rm Ia}$, therefore,
the initial mass of the WD, $M_{\rm WD,0} \simeq M_{\rm CO}$, should
be larger.  This implies that the older system should have larger
minimum $M_{\rm CO}$ as indicated in Figure \ref{umefig2}.  Using the
$X$(C)-$M_{\rm CO}$ and $M_{\rm Ni56}$-$X$(C) relations
(Fig. \ref{umefig2}), we conclude that WDs in older progenitor systems
have a smaller $X$(C), and thus produce dimmer SNe Ia.

\vspace*{-3mm}
\subsection{Comparison with observations}
\vspace*{-3mm}

 The first observational indication which can be compared with our
model is the possible dependence of the SN brightness on the
morphology of the host galaxies.  \cite{ham96} found that the
most luminous SNe Ia occur in spiral galaxies, while both spiral and
elliptical galaxies are hosts to dimmer SNe Ia. Hence, the mean peak
brightness is lower in elliptical than in spiral galaxies.

 In our model, this property is simply understood as the effect of the
different age of the companion. In spiral galaxies, star formation
occurs continuously up to the present time. Hence, both WD+MS and
WD+RG systems can produce SNe Ia. In elliptical galaxies, on the other
hand, star formation has long ended, typically more than 10 Gyr
ago. Hence, WD+MS systems can no longer produce SNe Ia. In Figure
\ref{umefig3}, we show the frequency of the expected SN I for a galaxy
of mass $2 \times 10^{11} M_\odot$ for WD+MS and WD+RG systems
separately as a function of $M_{\rm CO}$. Here we use the results of
\cite{hac99b,hac99a}, and the $M_{\rm CO}-X$(C) and $M_{\rm Ni56}- X$(C)
relations given in Figure \ref{umefig2}.  Since a WD with smaller
$M_{\rm CO}$ is assumed to produce a brighter SN Ia (larger $M_{\rm Ni
56}$), our model predicts that dimmer SNe Ia occur both in spirals and
in ellipticals, while brighter ones occur only in spirals.  The mean
brightness is smaller for ellipticals and the total SN Ia rate per
unit luminosity is larger in spirals than in ellipticals.  These
properties are consistent with observations.

 The second observational suggestion is the radial distribution of SNe
Ia in galaxies. \cite{wan97,rie98} found that
the variation of the peak brightness for SNe Ia located in the outer
regions in galaxies is smaller. This behavior can be understood as the
effect of metallicity.  As shown in Figure \ref{umefig2}, even when
the progenitor age is the same, the minimum $M_{\rm CO}$ is larger for
a smaller metallicity because of the metallicity dependence of the WD
winds. Therefore, our model predicts that the maximum brightness of
SNe Ia decreases as metallicity decreases.  Since the outer regions of
galaxies are thought to have lower metallicities than the inner
regions \cite{zar94,kob99}, our model
is consistent with observations. \cite{wan97} also claimed that
SNe Ia may be deficient in the bulges of spiral galaxies. This can be
explained by the age effect, because the bulge consists of old
population stars.

\vspace*{-3mm}
\subsection{Evolution of SNe Ia at high redshift}
\vspace*{-3mm}

 We have suggested that $X$(C) is the quantity very likely to cause
the diversity in $M_{\rm Ni56}$ and thus in the brightness of SNe Ia.
We have then shown that our model predicts that the brightness of SNe
Ia depends on the environment, in a way which is qualitatively
consistent with the observations. Further studies of the propagation
of the turbulent flame and the DDT are necessary in order to actually
prove that $X$(C) is the key parameter.

\begin{figure}
\centerline{\psfig{figure=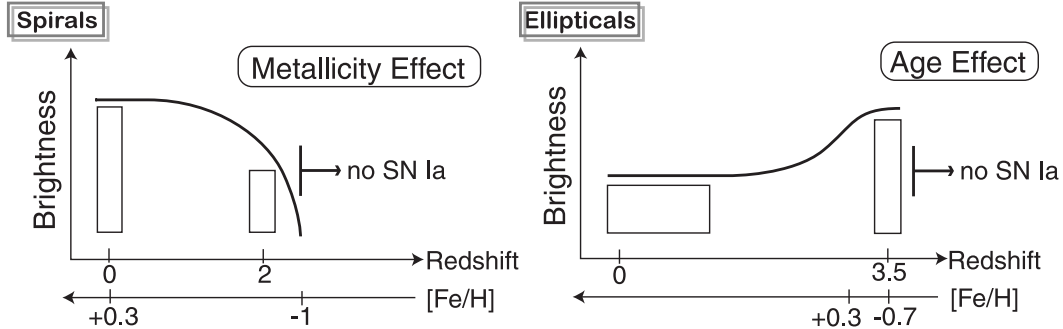,width=14cm}}
\caption{\label{sniadisp}
Illustration of the predicted variation in SN Ia brightness with
redshift.  }
\end{figure}

 Our model predicts that when the progenitors belong to an old
population, or to a low metal environment, the number of very bright
SNe Ia is small, so that the variation in brightness is also smaller,
which is shown in Figure \ref{sniadisp}. In spiral galaxies, the
metallicity is significantly smaller at redshifts $z\gsim 1$, and thus
both the mean brightness of SNe Ia and its range tend to be smaller
(Fig. \ref{sniadisp}).  At $z\gsim 2$ SNe Ia would not occur in
spirals at all because the metallicity is too low.  In elliptical
galaxies, on the other hand, the metallicity at redshifts $z \sim 1-3$
is not very different from the present value.  However, the age of the
galaxies at $z\simeq 1$ is only about 5 Gyr, so that the mean
brightness of SNe Ia and its range tend to be larger at $z\gsim 1$
than in the present ellipticals because of the age effect.

 We note that the variation of $X$(C) is larger in metal-rich nearby
spirals than in high redshift galaxies.  Therefore, if $X$(C) is the
main parameter responsible for the diversity of SNe Ia, and if the
LCS method is confirmed by the nearby SNe Ia data,
the LCS method can also be used to determine the absolute magnitude of
high redshift SNe Ia.

%%% fig 7

\vspace*{-3mm}
\subsection{Possible evolutionary effects}
\vspace*{-3mm}

In the above subsections, we consider the metallicity effects only on
the C/O ratio; this is just to shift the main-sequence mass - $M_{\rm
WD,0}$ relation, thus resulting in no important evolutionary effect.
However, some other metallicity effects could give rise to 
evolution of SNe Ia between high and low redshifts (i.e., between
low and high metallicities).

Here we point out just one possible metallicity effect on the carbon
ignition density in the accreting WD.  The ignition density is
determined by the competition between the compressional heating due to
accretion and the neutrino cooling.  The neutrino emission is enhanced
by the {\sl local} Urca shell process of, e.g., $^{21}$Ne--$^{21}$F
pair \cite{pac73}.  (Note that this is different from the {\sl
convective} Urca neutrino process).  For higher metallicity, the
abundance of $^{21}$Ne is larger so that the cooling is larger.  This
could delay the carbon ignition until a higher central density is
reached \cite{nom97d}.

Since the WD with a higher central density has a larger binding
energy, the kinetic energy of SNe Ia tends to be smaller if the same
amount of $^{56}$Ni is produced.  This might cause a systematically
slower light curve evolution at higher metallicity environment.  The
carbon ignition process including these metallicity effects as well as
the convective Urca neutrino process need to be studied (see also
\cite{iwa99} for nucleosynthesis constraints on the ignition
density).

\vspace*{-1mm}
\section{Cosmic supernova rates}
\vspace*{-1mm}

Attempts have been made to predict the cosmic supernova rates as a
function of redshift by using the observed cosmic star formation rate
(SFR) \cite{rui98,sad98,yun98}.  The observed cosmic
SFR shows a peak at $z \sim 1.4$ and a sharp decrease to the present
\cite{mad96}. However, UV luminosities which is converted to
the SFRs may be affected by the dust extinction \cite{pet98}.
Recent updates of the cosmic SFR suggest that a peak lies around $z
\sim 3$.

\cite{kob98} predicts that the cosmic SN Ia rate drops at
$z \sim 1-2$, due to the metallicity-dependence of the SN Ia rate.  
Their finding that the occurrence of SNe Ia depends on the metallicity
of the progenitor systems implies that the SN Ia rate strongly depends
on the history of the star formation and metal-enrichment.
The universe is composed of different morphological types of galaxies
and therefore the cosmic SFR is a sum of the SFRs for different types
of galaxies.  As each morphological type has a unique star formation
history, we should decompose the cosmic SFR into the SFR belonging to
each type of galaxy and calculate the SN Ia rate for each type of
galaxy.

Here we first construct the detailed evolution models for different
type of galaxies which are compatible with the stringent observational
constraints, and apply them to reproduce the cosmic SFR for two
different environments, e.g., the cluster and the field.  Secondly
with the self-consistent galaxy models, we calculate the SN rate
history for each type of galaxy and predict the cosmic supernova rates
as a function of redshift.

%-----------------------------------------------------------------

\vspace*{-3mm}
\subsection{In Clusters}
\vspace*{-3mm}

%%% fig 8
\begin{figure}
\centerline{\psfig{figure=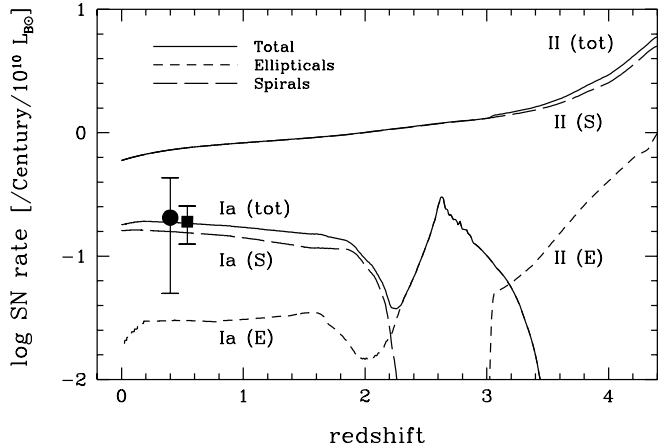,width=9.5cm}}
\caption[h]{\label{csnr_cluster}
The cosmic supernova rates (solid line) as the composite of ellipticals (short-dashed line) and spirals (long-dashed line).
The upper three lines show SN II rates, 
the lower three lines show SN Ia rates.
Observational data sources: 
circle, \cite{pai96}; square, \cite{pai99}.}
\end{figure}

Galaxies that are responsible for the cosmic SFR have different
timescales for the heavy-element enrichment, and the occurrence of
supernovae depends on the metallicity therein. Therefore
we calculate the cosmic supernova rate by summing up the supernova 
rates in spirals (S0a-Sa, Sab-Sb, Sbc-Sc, and Scd-Sd)
and ellipticals with the ratio of the relative mass
contribution.
The relative mass contribution is obtained from
the observed relative luminosity proportion 
and the calculated mass to light ratio in B-band \cite{kob00}.
The photometric evolution is calculated with the spectral synthesis
population database taken from \cite{kod97}. 
We adopt $H_0=50$ km s$^{-1}$ Mpc$^{-1}$, $\Omega_0=0.2$, $\lambda_0=0$, and
the galactic age of $t_{\rm age}=15$ Gyr.

First, we make a prediction of the cosmic supernova rates
by using the galaxy models
which are constructed to meet the observational constraints
of cluster galaxies.
We assume that elliptical galaxies are
formed by a single star burst and stop the star formation at $t \sim 1$ Gyr 
due to a supernova-driven galactic wind, 
while spiral galaxies are formed by a relatively continuous star formation.
The infall rates and the SFRs are given by \cite{kob00}.
These models are constructed to meet the latest observational
constraints such as the present gas fractions and colors for spirals,
and the mean stellar metallicity and the color evolution from the
present to $z \sim 1$ for ellipticals
(see \cite{kob00} for the figures).

The synthesized cosmic SFR has an excess
at $z \gtsim 3$ due to the early star burst in ellipticals and a
shallower slope from the present to the peak at $z \sim 1.4$, compared
with Madau's plot \cite{mad96}.
Figure \ref{csnr_cluster} shows the cosmic supernova rates in cluster galaxies.
The SN Ia rate in spirals drops at $z \sim 1.9$ 
because of the low-metallicity inhibition of SNe Ia. 
We can test the metallicity effect 
by finding this drop of the SN Ia in spirals,
if high-redshift SNe Ia at $z \gtsim 1.5$ and their host galaxies
are observed with the Next Generation Space Telescope.
In ellipticals, the chemical enrichment takes place so early that 
the metallicity is large enough to produce SNe Ia at $z \gtsim 2$. 
The two peaks of SN Ia rates at $z \sim 2.6$ and $z \sim 1.6$ 
come from the MS+WD and the RG+WD systems, respectively.
The SN Ia rate in ellipticals decreases at $z \sim 2.6$,
which is determined from the shortest lifetime of SNe Ia of $\sim 0.5$ Gyr.
Thus, the total SN Ia rate decrease at the same redshift as ellipticals, i.e.,
$z \sim 2.6$.
(Note, the decrease of the SN Ia rate at $z \sim 1.6$
disappears if we adopt $z_{\rm f}\sim 3$,
because the peak from the MS+WD systems moves to lower redshifts.)

\vspace*{-3mm}
\subsection{In Field}
\vspace*{-3mm}

%%% fig 9
\begin{figure}
\centerline{\psfig{figure=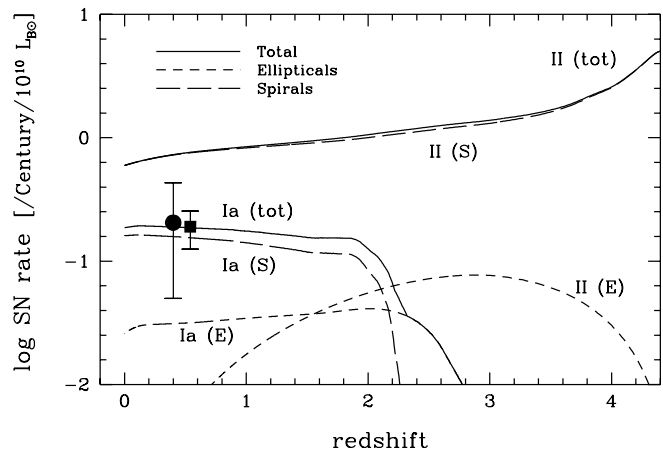,width=9.5cm}}
\caption[h]{\label{csnr_field}
The same as Figure \ref{csnr_cluster}, but for the formation epochs of ellipticals span at $1 \ltsim z \ltsim 4$, which is corresponding to field ellipticals.
}
\end{figure}

We also predict the cosmic supernova rates
assuming that the formation of ellipticals in field
took place for over the wide range of redshifts,
which is imprinted in the observed spectra of ellipticals
in the Hubble Deep Field \cite{fra98}.
The adopted SFRs are the same as the case of cluster galaxies,
but for the formation epochs $z_{\rm f}$ of ellipticals
distribute as $\propto\exp(-((z-2)/2)^2)$ in the range of $0 \le z \le 5$,

The synthesized
cosmic SFR has a broad peak around $z \sim 3$, which is in good
agreement with the recent sub-mm observation \cite{hug98}.
Figure \ref{csnr_field} shows the cosmic supernova rates in field galaxies.
As in Figure \ref{csnr_cluster},
the SN Ia rate in spirals drops at $z \sim 1.9$. 
The averaged SN Ia rate in ellipticals decreases at $z \sim 2.2$ 
as a result of $\sim 0.5$ Gyr delay of the decrease in the SFR at $z \gtsim 3$.
Then, the total SN Ia rate decreases gradually
from $z \sim 2$ to $z \sim 3$.

The rate of SNe II in ellipticals evolves 
following the SFR without time delay.
Then, it is possible to observe SNe II in ellipticals around $z \sim 1$.
The difference in the SN II and Ia rates between
cluster and field ellipticals reflects the difference in the
galaxy formation histories in the different environments.

\vspace*{-6mm}
\subsection{Summary}
\vspace*{-4mm}

(1) In the cluster environment, 
the predicted cosmic supernova rate suggests that
in ellipticals SNe Ia can be observed even at high redshifts because
the chemical enrichment takes place so early that the metallicity is
large enough to produce SNe Ia at $z \gtsim 2.5$. In spirals the SN Ia
rate drops at $z \sim 2$ because of the low-metallicity inhibition of
SNe Ia.

(2)In the field environment, ellipticals are assumed to form at such a
wide range of redshifts as $1 \ltsim z \ltsim 4$.   The SN
Ia rate is expected to be significantly low beyond $z \gtsim 2$ because
the SN Ia rate drops at $z \sim 2$ in spirals and gradually decreases
from $z \sim2 $ in ellipticals.

%-----------------------------------------------------------------
%\smallskip
%This work has been supported in part by the grant-in-Aid for COE
%Scientific Research (07CE2002) of the Ministry of Education, Science,
%and Culture in Japan.

\end{document}

%% file: pah0.tex
\def\PsfigVersion{1.10}
\def\setDriver{\DvipsDriver} % \DvipsDriver or \OzTeXDriver
\ifx\undefined\psfig\else \fi
\let\LaTeXAtSign=\@
\let\@=\relax
\edef\psfigRestoreAt{\catcode`\@=\number\catcode`@\relax}
\catcode`\@=11\relax
\newwrite\@unused
\def\ps@typeout#1{{\let\protect\string\immediate\write\@unused{#1}}}

\def\DvipsDriver{
	\ps@typeout{psfig/tex \PsfigVersion -dvips}
\def\PsfigSpecials{\DvipsSpecials} 	\def\ps@dir{/}
\def\ps@predir{} }
\def\OzTeXDriver{
	\ps@typeout{psfig/tex \PsfigVersion -oztex}
	\def\PsfigSpecials{\OzTeXSpecials}
	\def\ps@dir{:}
	\def\ps@predir{:}
	\catcode`\^^J=5
}

\def\figurepath{./:}

\def\DoPaths#1{\expandafter\EachPath#1\stoplist}
\def\leer{}
\def\EachPath#1:#2\stoplist{% #1 part of the list (delimiter :)
  \ExistsFile{#1}{\SearchedFile}
  \ifx#2\leer
  \else
    \expandafter\EachPath#2\stoplist
  \fi}
\def\ps@dir{/}
\def\ExistsFile#1#2{%
   \openin1=\ps@predir#1\ps@dir#2
   \ifeof1
       \closein1
   \else
       \closein1
        \ifx\ps@founddir\leer
           \edef\ps@founddir{#1}
        \fi
   \fi}
\def\get@dir#1{%
  \def\ps@founddir{}
  \def\SearchedFile{#1}
  \DoPaths\figurepath
}
\def\@nnil{\@nil}
\def\@empty{}
\def\@psdonoop#1\@@#2#3{}
\def\@psdo#1:=#2\do#3{\edef\@psdotmp{#2}\ifx\@psdotmp\@empty \else
    \expandafter\@psdoloop#2,\@nil,\@nil\@@#1{#3}\fi}
\def\@psdoloop#1,#2,#3\@@#4#5{\def#4{#1}\ifx #4\@nnil \else
       #5\def#4{#2}\ifx #4\@nnil \else#5\@ipsdoloop #3\@@#4{#5}\fi\fi}
\def\@ipsdoloop#1,#2\@@#3#4{\def#3{#1}\ifx #3\@nnil 
       \let\@nextwhile=\@psdonoop \else
      #4\relax\let\@nextwhile=\@ipsdoloop\fi\@nextwhile#2\@@#3{#4}}
\def\@tpsdo#1:=#2\do#3{\xdef\@psdotmp{#2}\ifx\@psdotmp\@empty \else
    \@tpsdoloop#2\@nil\@nil\@@#1{#3}\fi}
\def\@tpsdoloop#1#2\@@#3#4{\def#3{#1}\ifx #3\@nnil 
       \let\@nextwhile=\@psdonoop \else
      #4\relax\let\@nextwhile=\@tpsdoloop\fi\@nextwhile#2\@@#3{#4}}
\ifx\undefined\fbox
\newdimen\fboxrule
\newdimen\fboxsep
\newdimen\ps@tempdima
\newbox\ps@tempboxa
\long\def\fbox#1{\leavevmode\setbox\ps@tempboxa\hbox{#1}\ps@tempdima\fboxrule
    \advance\ps@tempdima \fboxsep \advance\ps@tempdima \dp\ps@tempboxa
   \hbox{\lower \ps@tempdima\hbox
  {\vbox{\hrule height \fboxrule
          \hbox{\vrule width \fboxrule \hskip\fboxsep
          \vbox{\vskip\fboxsep \box\ps@tempboxa\vskip\fboxsep}\hskip 
                 \fboxsep\vrule width \fboxrule}
                 \hrule height \fboxrule}}}}
\fi
\newread\ps@stream
\newif\ifnot@eof       % continue looking for the bounding box?
\newif\if@noisy        % report what you're making?
\newif\if@atend        % %%BoundingBox: has (at end) specification
\newif\if@psfile       % does this look like a PostScript file?
{\catcode`\%=12\global\gdef\epsf@start{%!}}
\def\epsf@PS{PS}
\def\epsf@getbb#1{%
\openin\ps@stream=\ps@predir#1
\ifeof\ps@stream\ps@typeout{Error, File #1 not found}\else
   {\not@eoftrue \chardef\other=12
    \def\do##1{\catcode`##1=\other}\dospecials \catcode`\ =10
    \loop
       \if@psfile
	  \read\ps@stream to \epsf@fileline
       \else{
	  \obeyspaces
          \read\ps@stream to \epsf@tmp\global\let\epsf@fileline\epsf@tmp}
       \fi
       \ifeof\ps@stream\not@eoffalse\else
       \if@psfile\else
       \expandafter\epsf@test\epsf@fileline:. \\%
       \fi
          \expandafter\epsf@aux\epsf@fileline:. \\%
       \fi
   \ifnot@eof\repeat
   }\closein\ps@stream\fi}%
\long\def\epsf@test#1#2#3:#4\\{\def\epsf@testit{#1#2}
			\ifx\epsf@testit\epsf@start\else
\ps@typeout{Warning! File does not start with `\epsf@start'.  It may not be a PostScript file.}
			\fi
			\@psfiletrue} % don't test after 1st line
{\catcode`\%=12\global\let\epsf@percent=%\global\def\epsf@bblit{%BoundingBox}}
\long\def\epsf@aux#1#2:#3\\{\ifx#1\epsf@percent
   \def\epsf@testit{#2}\ifx\epsf@testit\epsf@bblit
	\@atendfalse
        \epsf@atend #3 . \\%
	\if@atend	
	   \if@verbose{
		\ps@typeout{psfig: found `(atend)'; continuing search}
	   }\fi
        \else
        \epsf@grab #3 . . . \\%
        \not@eoffalse
        \global\no@bbfalse
        \fi
   \fi\fi}%
\def\epsf@grab #1 #2 #3 #4 #5\\{%
   \global\def\epsf@llx{#1}\ifx\epsf@llx\empty
      \epsf@grab #2 #3 #4 #5 .\\\else
   \global\def\epsf@lly{#2}%
   \global\def\epsf@urx{#3}\global\def\epsf@ury{#4}\fi}%
\def\epsf@atendlit{(atend)} 
\def\epsf@atend #1 #2 #3\\{%
   \def\epsf@tmp{#1}\ifx\epsf@tmp\empty
      \epsf@atend #2 #3 .\\\else
   \ifx\epsf@tmp\epsf@atendlit\@atendtrue\fi\fi}

\chardef\psletter = 11 % won't conflict with \begin{letter} now...
\chardef\other = 12

\newif \ifdebug %%% turn me on to see TeX hard at work ...
\newif\ifc@mpute %%% don't need to compute some values
\c@mputetrue % but assume that we do

\let\then = \relax
\def\r@dian{pt }
\let\r@dians = \r@dian
\let\dimensionless@nit = \r@dian
\let\dimensionless@nits = \dimensionless@nit
\def\internal@nit{sp }
\let\internal@nits = \internal@nit
\newif\ifstillc@nverging
\def \Mess@ge #1{\ifdebug \then \message {#1} \fi}

{ %%% Things that need abnormal catcodes %%%
	\catcode `\@ = \psletter
	\gdef \nodimen {\expandafter \n@dimen \the \dimen}
	\gdef \term #1 #2 #3%
	       {\edef \t@ {\the #1}%%% freeze parameter 1 (count, by value)
		\edef \t@@ {\expandafter \n@dimen \the #2\r@dian}%
		\t@rm {\t@} {\t@@} {#3}%
	       }
	\gdef \t@rm #1 #2 #3%
	       {{%
		\count 0 = 0
		\dimen 0 = 1 \dimensionless@nit
		\dimen 2 = #2\relax
		\Mess@ge {Calculating term #1 of \nodimen 2}%
		\loop
		\ifnum	\count 0 < #1
		\then	\advance \count 0 by 1
			\Mess@ge {Iteration \the \count 0 \space}%
			\Multiply \dimen 0 by {\dimen 2}%
			\Mess@ge {After multiplication, term = \nodimen 0}%
			\Divide \dimen 0 by {\count 0}%
			\Mess@ge {After division, term = \nodimen 0}%
		\repeat
		\Mess@ge {Final value for term #1 of 
				\nodimen 2 \space is \nodimen 0}%
		\xdef \Term {#3 = \nodimen 0 \r@dians}%
		\aftergroup \Term
	       }}
	\catcode `\p = \other
	\catcode `\t = \other
	\gdef \n@dimen #1pt{#1} %%% throw away the ``pt''
}

\def \Divide #1by #2{\divide #1 by #2} %%% just a synonym

\def \Multiply #1by #2%%% allows division of a dimen by a dimen
       {{%%% should really freeze parameter 2 (dimen, passed by value)
	\count 0 = #1\relax
	\count 2 = #2\relax
	\count 4 = 65536
	\Mess@ge {Before scaling, count 0 = \the \count 0 \space and
			count 2 = \the \count 2}%
	\ifnum	\count 0 > 32767 %%% do our best to avoid overflow
	\then	\divide \count 0 by 4
		\divide \count 4 by 4
	\else	\ifnum	\count 0 < -32767
		\then	\divide \count 0 by 4
			\divide \count 4 by 4
		\else
		\fi
	\fi
	\ifnum	\count 2 > 32767 %%% while retaining reasonable accuracy
	\then	\divide \count 2 by 4
		\divide \count 4 by 4
	\else	\ifnum	\count 2 < -32767
		\then	\divide \count 2 by 4
			\divide \count 4 by 4
		\else
		\fi
	\fi
	\multiply \count 0 by \count 2
	\divide \count 0 by \count 4
	\xdef \product {#1 = \the \count 0 \internal@nits}%
	\aftergroup \product
       }}

\def\r@duce{\ifdim\dimen0 > 90\r@dian \then   % sin(x+90) = sin(180-x)
		\multiply\dimen0 by -1
		\advance\dimen0 by 180\r@dian
		\r@duce
	    \else \ifdim\dimen0 < -90\r@dian \then  % sin(-x) = sin(360+x)
		\advance\dimen0 by 360\r@dian
		\r@duce
		\fi
	    \fi}

\def\Sine#1%
       {{%
	\dimen 0 = #1 \r@dian
	\r@duce
	\ifdim\dimen0 = -90\r@dian \then
	   \dimen4 = -1\r@dian
	   \c@mputefalse
	\fi
	\ifdim\dimen0 = 90\r@dian \then
	   \dimen4 = 1\r@dian
	   \c@mputefalse
	\fi
	\ifdim\dimen0 = 0\r@dian \then
	   \dimen4 = 0\r@dian
	   \c@mputefalse
	\fi
	\ifc@mpute \then
		\divide\dimen0 by 180
		\dimen0=3.141592654\dimen0
		\dimen 2 = 3.1415926535897963\r@dian %%% a well-known constant
		\divide\dimen 2 by 2 %%% we only deal with -pi/2 : pi/2
		\Mess@ge {Sin: calculating Sin of \nodimen 0}%
		\count 0 = 1 %%% see power-series expansion for sine
		\dimen 2 = 1 \r@dian %%% ditto
		\dimen 4 = 0 \r@dian %%% ditto
		\loop
			\ifnum	\dimen 2 = 0 %%% then we've done
			\then	\stillc@nvergingfalse 
			\else	\stillc@nvergingtrue
			\fi
			\ifstillc@nverging %%% then calculate next term
			\then	\term {\count 0} {\dimen 0} {\dimen 2}%
				\advance \count 0 by 2
				\count 2 = \count 0
				\divide \count 2 by 2
				\ifodd	\count 2 %%% signs alternate
				\then	\advance \dimen 4 by \dimen 2
				\else	\advance \dimen 4 by -\dimen 2
				\fi
		\repeat
	\fi		
			\xdef \sine {\nodimen 4}%
       }}

\def\Cosine#1{\ifx\sine\UnDefined\edef\Savesine{\relax}\else
		             \edef\Savesine{\sine}\fi
	{\dimen0=#1\r@dian\advance\dimen0 by 90\r@dian
	 \Sine{\nodimen 0}
	 \xdef\cosine{\sine}
	 \xdef\sine{\Savesine}}}	      
\def\psdraft{
	\def\@psdraft{0}
}
\def\psfull{
	\def\@psdraft{100}
}

\psfull

\newif\if@scalefirst
\def\psscalefirst{\@scalefirsttrue}
\def\psrotatefirst{\@scalefirstfalse}
\psrotatefirst

\newif\if@draftbox
\def\psnodraftbox{
	\@draftboxfalse
}
\def\psdraftbox{
	\@draftboxtrue
}
\@draftboxtrue

\newif\if@prologfile
\newif\if@postlogfile
\def\pssilent{
	\@noisyfalse
}
\def\psnoisy{
	\@noisytrue
}
\psnoisy
\newif\if@bbllx
\newif\if@bblly
\newif\if@bburx
\newif\if@bbury
\newif\if@height
\newif\if@width
\newif\if@rheight
\newif\if@rwidth
\newif\if@angle
\newif\if@clip
\newif\if@verbose
\def\@p@@sclip#1{\@cliptrue}
\newif\if@decmpr
\def\@p@@sfigure#1{\def\@p@sfile{null}\def\@p@sbbfile{null}\@decmprfalse
   \openin1=\ps@predir#1
   \ifeof1
	\closein1
	\get@dir{#1}
	\ifx\ps@founddir\leer
		\openin1=\ps@predir#1.bb
		\ifeof1
			\closein1
			\get@dir{#1.bb}
			\ifx\ps@founddir\leer
				\ps@typeout{Can't find #1 in \figurepath}
			\else
				\@decmprtrue
				\def\@p@sfile{\ps@founddir\ps@dir#1}
				\def\@p@sbbfile{\ps@founddir\ps@dir#1.bb}
			\fi
		\else
			\closein1
			\@decmprtrue
			\def\@p@sfile{#1}
			\def\@p@sbbfile{#1.bb}
		\fi
	\else
		\def\@p@sfile{\ps@founddir\ps@dir#1}
		\def\@p@sbbfile{\ps@founddir\ps@dir#1}
	\fi
   \else
	\closein1
	\def\@p@sfile{#1}
	\def\@p@sbbfile{#1}
   \fi
}
\def\@p@@sfile#1{\@p@@sfigure{#1}}
\def\@p@@sbbllx#1{
		\@bbllxtrue
		\dimen100=#1
		\edef\@p@sbbllx{\number\dimen100}
}
\def\@p@@sbblly#1{
		\@bbllytrue
		\dimen100=#1
		\edef\@p@sbblly{\number\dimen100}
}
\def\@p@@sbburx#1{
		\@bburxtrue
		\dimen100=#1
		\edef\@p@sbburx{\number\dimen100}
}
\def\@p@@sbbury#1{
		\@bburytrue
		\dimen100=#1
		\edef\@p@sbbury{\number\dimen100}
}
\def\@p@@sheight#1{
		\@heighttrue
		\dimen100=#1
   		\edef\@p@sheight{\number\dimen100}
}
\def\@p@@swidth#1{
		\@widthtrue
		\dimen100=#1
		\edef\@p@swidth{\number\dimen100}
}
\def\@p@@srheight#1{
		\@rheighttrue
		\dimen100=#1
		\edef\@p@srheight{\number\dimen100}
}
\def\@p@@srwidth#1{
		\@rwidthtrue
		\dimen100=#1
		\edef\@p@srwidth{\number\dimen100}
}
\def\@p@@sangle#1{
		\@angletrue
		\edef\@p@sangle{#1} %\number\dimen100}
}
\def\@p@@ssilent#1{ 
		\@verbosefalse
}
\def\@p@@sprolog#1{\@prologfiletrue\def\@prologfileval{#1}}
\def\@p@@spostlog#1{\@postlogfiletrue\def\@postlogfileval{#1}}
\def\@cs@name#1{\csname #1\endcsname}
\def\@setparms#1=#2,{\@cs@name{@p@@s#1}{#2}}
\def\ps@init@parms{
		\@bbllxfalse \@bbllyfalse
		\@bburxfalse \@bburyfalse
		\@heightfalse \@widthfalse
		\@rheightfalse \@rwidthfalse
		\def\@p@sbbllx{}\def\@p@sbblly{}
		\def\@p@sbburx{}\def\@p@sbbury{}
		\def\@p@sheight{}\def\@p@swidth{}
		\def\@p@srheight{}\def\@p@srwidth{}
		\def\@p@sangle{0}
		\def\@p@sfile{} \def\@p@sbbfile{}
		\def\@p@scost{10}
		\def\@sc{}
		\@prologfilefalse
		\@postlogfilefalse
		\@clipfalse
		\if@noisy
			\@verbosetrue
		\else
			\@verbosefalse
		\fi
}
\def\parse@ps@parms#1{
	 	\@psdo\@psfiga:=#1\do
		   {\expandafter\@setparms\@psfiga,}}
\newif\ifno@bb
\def\bb@missing{
	\if@verbose{
		\ps@typeout{psfig: searching \@p@sbbfile \space  for bounding box}
	}\fi
	\no@bbtrue
	\epsf@getbb{\@p@sbbfile}
        \ifno@bb \else \bb@cull\epsf@llx\epsf@lly\epsf@urx\epsf@ury\fi
}	
\def\bb@cull#1#2#3#4{
	\dimen100=#1 bp\edef\@p@sbbllx{\number\dimen100}
	\dimen100=#2 bp\edef\@p@sbblly{\number\dimen100}
	\dimen100=#3 bp\edef\@p@sbburx{\number\dimen100}
	\dimen100=#4 bp\edef\@p@sbbury{\number\dimen100}
	\no@bbfalse
}
\newdimen\p@intvaluex
\newdimen\p@intvaluey
\def\rotate@#1#2{{\dimen0=#1 sp\dimen1=#2 sp
		  \global\p@intvaluex=\cosine\dimen0
		  \dimen3=\sine\dimen1
		  \global\advance\p@intvaluex by -\dimen3
		  \global\p@intvaluey=\sine\dimen0
		  \dimen3=\cosine\dimen1
		  \global\advance\p@intvaluey by \dimen3
		  }}
\def\compute@bb{
		\no@bbfalse
		\if@bbllx \else \no@bbtrue \fi
		\if@bblly \else \no@bbtrue \fi
		\if@bburx \else \no@bbtrue \fi
		\if@bbury \else \no@bbtrue \fi
		\ifno@bb \bb@missing \fi
		\ifno@bb \ps@typeout{FATAL ERROR: no bb supplied or found}
			\no-bb-error
		\fi
		\count203=\@p@sbburx
		\count204=\@p@sbbury
		\advance\count203 by -\@p@sbbllx
		\advance\count204 by -\@p@sbblly
		\edef\ps@bbw{\number\count203}
		\edef\ps@bbh{\number\count204}
		\if@angle 
			\Sine{\@p@sangle}\Cosine{\@p@sangle}
	        	{\dimen100=\maxdimen\xdef\r@p@sbbllx{\number\dimen100}
					    \xdef\r@p@sbblly{\number\dimen100}
			                    \xdef\r@p@sbburx{-\number\dimen100}
					    \xdef\r@p@sbbury{-\number\dimen100}}
                        \def\minmaxtest{
			   \ifnum\number\p@intvaluex<\r@p@sbbllx
			      \xdef\r@p@sbbllx{\number\p@intvaluex}\fi
			   \ifnum\number\p@intvaluex>\r@p@sbburx
			      \xdef\r@p@sbburx{\number\p@intvaluex}\fi
			   \ifnum\number\p@intvaluey<\r@p@sbblly
			      \xdef\r@p@sbblly{\number\p@intvaluey}\fi
			   \ifnum\number\p@intvaluey>\r@p@sbbury
			      \xdef\r@p@sbbury{\number\p@intvaluey}\fi
			   }
			\rotate@{\@p@sbbllx}{\@p@sbblly}
			\minmaxtest
			\rotate@{\@p@sbbllx}{\@p@sbbury}
			\minmaxtest
			\rotate@{\@p@sbburx}{\@p@sbblly}
			\minmaxtest
			\rotate@{\@p@sbburx}{\@p@sbbury}
			\minmaxtest
			\edef\@p@sbbllx{\r@p@sbbllx}\edef\@p@sbblly{\r@p@sbblly}
			\edef\@p@sbburx{\r@p@sbburx}\edef\@p@sbbury{\r@p@sbbury}
		\fi
		\count203=\@p@sbburx
		\count204=\@p@sbbury
		\advance\count203 by -\@p@sbbllx
		\advance\count204 by -\@p@sbblly
		\edef\@bbw{\number\count203}
		\edef\@bbh{\number\count204}
}
\def\in@hundreds#1#2#3{\count240=#2 \count241=#3
		     \count100=\count240	% 100 is first digit #2/#3
		     \divide\count100 by \count241
		     \count101=\count100
		     \multiply\count101 by \count241
		     \advance\count240 by -\count101
		     \multiply\count240 by 10
		     \count101=\count240	%101 is second digit of #2/#3
		     \divide\count101 by \count241
		     \count102=\count101
		     \multiply\count102 by \count241
		     \advance\count240 by -\count102
		     \multiply\count240 by 10
		     \count102=\count240	% 102 is the third digit
		     \divide\count102 by \count241
		     \count200=#1\count205=0
		     \count201=\count200
			\multiply\count201 by \count100
		 	\advance\count205 by \count201
		     \count201=\count200
			\divide\count201 by 10
			\multiply\count201 by \count101
			\advance\count205 by \count201
		     \count201=\count200
			\divide\count201 by 100
			\multiply\count201 by \count102
			\advance\count205 by \count201
		     \edef\@result{\number\count205}
}
\def\compute@wfromh{
		\in@hundreds{\@p@sheight}{\@bbw}{\@bbh}
		\edef\@p@swidth{\@result}
}
\def\compute@hfromw{
	        \in@hundreds{\@p@swidth}{\@bbh}{\@bbw}
		\edef\@p@sheight{\@result}
}
\def\compute@handw{
		\if@height 
			\if@width
			\else
				\compute@wfromh
			\fi
		\else 
			\if@width
				\compute@hfromw
			\else
				\edef\@p@sheight{\@bbh}
				\edef\@p@swidth{\@bbw}
			\fi
		\fi
}
\def\compute@resv{
		\if@rheight \else \edef\@p@srheight{\@p@sheight} \fi
		\if@rwidth \else \edef\@p@srwidth{\@p@swidth} \fi
}
\def\compute@sizes{
	\compute@bb
	\if@scalefirst\if@angle
	\if@width
	   \in@hundreds{\@p@swidth}{\@bbw}{\ps@bbw}
	   \edef\@p@swidth{\@result}
	\fi
	\if@height
	   \in@hundreds{\@p@sheight}{\@bbh}{\ps@bbh}
	   \edef\@p@sheight{\@result}
	\fi
	\fi\fi
	\compute@handw
	\compute@resv}
\def\OzTeXSpecials{
	\special{empty.ps /@isp {true} def}
	\special{empty.ps \@p@swidth \space \@p@sheight \space
			\@p@sbbllx \space \@p@sbblly \space
			\@p@sbburx \space \@p@sbbury \space
			startTexFig \space }
	\if@clip{
		\if@verbose{
			\ps@typeout{(clip)}
		}\fi
		\special{empty.ps doclip \space }
	}\fi
	\if@angle{
		\if@verbose{
			\ps@typeout{(rotate)}
		}\fi
		\special {empty.ps \@p@sangle \space rotate \space} 
	}\fi
	\if@prologfile
	    \special{\@prologfileval \space } \fi
	\if@decmpr{
		\if@verbose{
			\ps@typeout{psfig: Compression not available
			in OzTeX version \space }
		}\fi
	}\else{
		\if@verbose{
			\ps@typeout{psfig: including \@p@sfile \space }
		}\fi
		\special{epsf=\ps@predir\@p@sfile \space }
	}\fi
	\if@postlogfile
	    \special{\@postlogfileval \space } \fi
	\special{empty.ps /@isp {false} def}
}
\def\DvipsSpecials{
	\special{ps::[begin] 	\@p@swidth \space \@p@sheight \space
			\@p@sbbllx \space \@p@sbblly \space
			\@p@sbburx \space \@p@sbbury \space
			startTexFig \space }
	\if@clip{
		\if@verbose{
			\ps@typeout{(clip)}
		}\fi
		\special{ps:: doclip \space }
	}\fi
	\if@angle
		\if@verbose{
			\ps@typeout{(clip)}
		}\fi
		\special {ps:: \@p@sangle \space rotate \space} 
	\fi
	\if@prologfile
	    \special{ps: plotfile \@prologfileval \space } \fi
	\if@decmpr{
		\if@verbose{
			\ps@typeout{psfig: including \@p@sfile.Z \space }
		}\fi
		\special{ps: plotfile "`zcat \@p@sfile.Z" \space }
	}\else{
		\if@verbose{
			\ps@typeout{psfig: including \@p@sfile \space }
		}\fi
		\special{ps: plotfile \@p@sfile \space }
	}\fi
	\if@postlogfile
	    \special{ps: plotfile \@postlogfileval \space } \fi
	\special{ps::[end] endTexFig \space }
}
\def\psfig#1{\vbox {
	\ps@init@parms
	\parse@ps@parms{#1}
	\compute@sizes
	\ifnum\@p@scost<\@psdraft{
		\PsfigSpecials 
		\vbox to \@p@srheight sp{
			\hbox to \@p@srwidth sp{
				\hss
			}
		\vss
		}
	}\else{
		\if@draftbox{		
			\hbox{\fbox{\vbox to \@p@srheight sp{
			\vss
			\hbox to \@p@srwidth sp{ \hss 
			 \hss }
			\vss
			}}}
		}\else{
			\vbox to \@p@srheight sp{
			\vss
			\hbox to \@p@srwidth sp{\hss}
			\vss
			}
		}\fi

	}\fi
}}
\psfigRestoreAt
\setDriver
\let\@=\LaTeXAtSign